\documentstyle[preprint,aps,epsfig]{revtex}
\begin{document}
\draft
\title{Field-induced structure transformation in electrorheological solids}
\author{C. K. Lo and K. W. Yu}
\address{Department of Physics, The Chinese University of Hong Kong, \\
         Shatin, New Territories, Hong Kong, China }
 
\maketitle

\begin{abstract}
We have computed the local electric field in a body-centered tetragonal 
(BCT) lattice of point dipoles via the Ewald-Kornfeld formulation, 
in an attempt to examine the effects of a structure transformation on 
the local field strength.
For the ground state of an electrorheological solid of hard spheres, 
we identified a novel structure transformation from the BCT to the 
face-centered cubic (FCC) lattices by changing the uniaxial lattice 
constant $c$ under the hard sphere constraint. 
In contrast to the previous results,
the local field exhibits a non-monotonic transition from BCT to FCC.
As $c$ increases from the BCT ground state, the local field initially 
decreases rapidly towards the isotropic value at the body-centered cubic 
lattice, decreases further, reaching a minimum value and increases, 
passing through the isotropic value again at an intermediate lattice, 
reaches a maximum value and finally decreases to the FCC value.
An experimental realization of the structure transformation is suggested.
Moreover, the change in the local field can lead to a generalized 
Clausius-Mossotti equation for the BCT lattices.
\end{abstract}
\vskip 5mm
\pacs{PACS Numbers: 83.80.Gv, 83.20.Di, 77.22.-d, 02.60.Lj}

\section{Introduction}

When a strong field is applied to a composite medium, 
the induced change of the medium can lead to spectacular
behavior, both in electrical transport and in optical response 
\cite{ETOPIM5}.
If a strong electric field is applied to a suspension of particles where
the particles have a large electric polarizability, the induced dipole 
moments of the particles can order the suspended particles into a 
body-centered tetragonal (BCT) lattice \cite{Tao}. 
This process is known as the electrorheological (ER) effect.

Recently, Tao and coworkers \cite{Tao98} proposed that a structure 
transformation from the BCT ground state to some other lattices can occur 
when one simultaneously applies a magnetic field perpendicular to the 
electric field and the polarized particles possess magnetic dipole 
moments. 
Sheng and coworkers \cite{Sheng} verified the proposal experimentally and 
observed a structure transformation from the BCT to face-centered cubic 
(FCC) lattices. Motivated by these studies, we propose an alternative 
structure transformation from the BCT to the FCC structure,
through the application of electric fields only.

The plan of the paper is as follows. 
We will adopt the point-dipole approximation \cite{Tao} and calculate the 
dipole lattice sum via the Ewald-Kornfeld formulation in section II.
In section III, we discuss the effects of a structure change on the local 
field when the lattice constants vary. 
If we change the uniaxial lattice constant under the hard-sphere 
condition, 
a series of transformations occur among the BCT ground state, 
body-centered cubic (BCC), intermediate and FCC lattices (see Fig.1). 
The results will be compared with those of a tetragonal lattice.
We also make a contact with macroscopic concept and derive 
the Clausius-Mossotti equation.
In section IV, we will compute the dipole interaction energy as 
a function of various lattices and discuss a possible structure 
transformation by the application of rotating electric fields.
Discussion and conclusion on our results will be given.

\section{Formalism}

In this section, we apply the Ewald-Kornfeld formulation 
\cite{Ewald,Kornfeld} 
to compute the local electric field for a BCT lattice of point dipoles. 
The BCT lattice can be regarded as a tetragonal lattice, 
plus a basis of two point dipoles, one of which is located at a corner 
and the other of which at the body center of the tetragonal unit cell. 
The tetragonal lattice has a lattice constant $c=q\xi$ along the z-axis 
and lattice constants $a=b=\xi q^{-1/2}$ along the x and y axes.
The volume of the tetragonal unit cell remains $V_c=\xi^3$ as $q$ varies.
In this way, the degree of anisotropy of the tetragonal lattice is 
measured 
by how $q$ is deviated from unity and the uniaxial anisotropic axis is 
along the z-axis. 

So far, the lattice parameter $\xi$ remains arbitrary. For hard spheres in
an ER solid, however, the lattice parameter $\xi$ can be determined from 
the relation: $2a^2+c^2=16 R^2$, where $R$ is the radius of the spheres. 
The hard sphere condition requires $a \ge 2R$ and $c \ge 2R$.
For the BCT ground state, $a=b=\sqrt{6}R$ and $c=2R$, while for the FCC
lattice, $a=b=2R$ and $c=\sqrt{8}R$ \cite{Tao}. The BCC lattice
is characterized by $a=b=c=\sqrt{16/3}R$.
Each sphere has a point dipole embedded at its center.

The lattice vector of the tetragonal lattice is given by
\begin{eqnarray}
{\bf R} = \xi(q^{-1/2} l\hat{\bf x} + q^{-1/2} m\hat{\bf y} + q n\hat{\bf 
z}),
\end{eqnarray}
where $l, m, n$ are integers. Suppose there are $N$ point dipoles ${\bf 
p}_i$ 
located at ${\bf r}_i$ in a unit cell. 
The local electric field ${\bf E}_i$ at a particular point dipole at 
${\bf r}_i$ can be expressed as a sum of the electric field of all 
dipoles 
at ${\bf r}_{{\bf R} j}$:
\begin{eqnarray}
{\bf E}_i = \sum'_j \sum_{\bf R} {\sf T}_{i {\bf R} j} \cdot 
  {\bf p}_j,
\label{dipole-sum}
\end{eqnarray}
where ``prime'' denotes a restricted summation which excludes $j=i$ when
${\bf R}=0$ and
\begin{eqnarray}
{\sf T}_{ij} = -\nabla_i \nabla_j {1\over |{\bf r}_i - {\bf r}_j|}
\end{eqnarray}
is the dipole interaction tensor.
Eq.(\ref{dipole-sum}) can be recast in the Ewald-Kornfeld form
\cite{Ewald,Kornfeld}:
\begin{eqnarray}
{\bf p}_i \cdot {\bf E}_i &=&
\sum'_j \sum_{\bf R} \left[-({\bf p}_i \cdot {\bf p}_j) B(r_{i {\bf R} j})
 + ({\bf p}_i \cdot {\bf r}_{i {\bf R} j})
   ({\bf p}_j \cdot {\bf r}_{i {\bf R} j}) C(r_{i {\bf R} j})\right] 
   \nonumber \\
 &-& {4\pi \over V_c} \sum_{{\bf G} \neq 0}
   {1\over G^2} \exp \left(-{G^2\over 4\eta^2} \right)
   [({\bf p}_i \cdot {\bf G}) \exp(i{\bf G} \cdot {\bf r}_i)
   \sum_j ({\bf p}_j \cdot {\bf G}) \exp(-i{\bf G} \cdot {\bf r}_j)]
   \nonumber \\
 &+& {4\eta^3 p_i^2 \over 3\sqrt{\pi}},
\label{E-K}
\end{eqnarray}
where $r_{i {\bf R} j}=|{\bf r}_i - {\bf r}_{{\bf R}j}|$, $\eta$ is 
an adjustable parameter and {\bf G} is a reciprocal lattice vector:
\begin{eqnarray}
{\bf G} = {2\pi\over \xi} (q^{1/2} u\hat{\bf x} + q^{1/2} v\hat{\bf y} 
 + q^{-1} w\hat{\bf z}).
\end{eqnarray}
The $B$ and $C$ coefficients are given by:
\begin{eqnarray}
B(r) &=& {{\rm erfc}(\eta r)\over r^3} + {2\eta \over \sqrt{\pi} r^2} 
  \exp(-\eta^2 r^2), \\
C(r) &=& {3 {\rm erfc}(\eta r)\over r^5} 
+\left({4\eta^3 \over \sqrt{\pi} r^2} + {6\eta \over \sqrt{\pi} r^4}\right)
   \exp(-\eta^2 r^2),
\end{eqnarray}
where ${\rm erfc}(r)$ is the complementary error function. 
Thus the dipole lattice sum of Eq.(\ref{dipole-sum}) becomes a summation 
over the real lattice vector {\bf R} as well as the reciprocal lattice 
vector {\bf G}.
Here we have considered an infinite lattice. For finite lattices, one 
must be careful about the effects of different boundary conditions
\cite{Allen}.

We should remark that although a tetragonal lattice is considered,
Eq.(\ref{E-K}) is applicable to arbitrary Bravais lattices.
The adjustable parameter $\eta$ is chosen so that both the summations 
in the real and reciprocal lattices converge most rapidly.
In what follows, we will limit ourselves to the BCT cell with two dipoles 
per tetragonal cell, 
and the Ewald-Kornfeld summation [Eq.(\ref{E-K})] can be carried out.
We will consider two cases depending on whether the dipole moment is 
parallel or perpendicular to the uniaxial anisotropic axis. 
In both cases, we will compute the local field as a function of the 
degree of anisotropy $q$.

\section{Effects of structure transformation on the local field}

Consider the longitudinal field case: ${\bf p}=p\hat{\bf z}$,
i.e., the dipole moments being along the uniaxial anisotropic axis.  
The local field {\bf E} at the the lattice point {\bf R} = 0 reduces to
\begin{eqnarray}
E_z = p \sum_{j=1}^{2} \sum'_{\bf R}[-B(R_j)+{z_j^2}{q^2} C(R_j)]
 - {4\pi p\over V_c} \sum_{{\bf G}\neq 0} S({\bf G}) {G_z^2\over G^2}
  \exp\left({-G^2\over 4\eta^2}\right) + {4p\eta^3\over {3\sqrt\pi}},
\end{eqnarray}
and $E_x=E_y=0$.
In the equation, $z_j$ and $R_j$ are respectively given by:
$$
z_j=n-{j-1\over 2},\ \ \ 
R_j=\left|{\bf R}-{j-1\over 2}(a\hat{\bf x}+a\hat{\bf y}+c\hat{\bf 
z})\right|,
$$
and $S({\bf G})=1+\exp[i(u+v+w)/\pi]$ is the structure factor.
The local field will be computed by summing over all integer indices,
$(j,l,m,n) \neq (1,0,0,0)$ for the summation in the real lattice and
$(u,v,w) \neq (0,0,0)$ for that in the reciprocal lattice.
Because of the exponential factors, we may impose an upper limit to the 
indices, i.e., all indices ranging from $-L$ to $L$,
where $L$ is a positive integer.
For $q \neq 1$, the regions of summation will be rectangular rather than
cubic in both the real and reciprocal lattices.
The computation has been repeated for various degree of anisotropy with
$q$ ranging from 0.5 to 2.0. A plateau value for $E_z$ is found for each
$q$ within a certain range of $\eta$ values: $1<\eta<10^{0.6}$. 
For instance, the calculations with $\eta=10^{0.5}$ yield numerical 
results already accurate up to 16 significant figures, 
indicating that convergence of the local field has indeed been achieved 
with the upper limit $L=4$.
For larger anisotropy (either $q \ll 1$ or $q \gg 1$), a spherical region
of summation can help the convergence \cite{Lo}. 

For the transverse field case in which the dipole moments are perpendicular
to the uniaxial anisotropic axis, Eq.(8) can still be applied to evaluate
the local field by modifying $G_z$ to $G_x$ while taking the gradient
along the direction of the dipole, say the x-axis,
and obtain the expression of the local field.

The results of the local field strength (normalized to $4\pi P/3$) against
$\log_{10} q$ for the longitudinal and transverse field cases are plotted 
in Fig.2(a) and Fig.2(b) respectively. For comparison, the corresponding 
results for a tetragonal lattice \cite{Lo} (i.e., in the absence of the 
body centers) are also plotted on the same figure.
As $q$ decreases, the local field for the longitudinal field case increases
rapidly while that for the transverse field case decreases rapidly.  
In both cases, when $q$ deviates from unity, the effect of anisotropy has 
a pronounced effect on the local field strengths.

Unlike the tetragonal case, the local field of the BCT lattice does not 
vary 
much near $q=1$. When we magnify the scales in Fig.2(c) and Fig.2(d), 
we observe a non-monotonic behavior as $q$ increases: 
as $c$ increases from BCT, the local field initially decreases rapidly
towards the isotropic value at BCC, decreases further,
reaching a minimum value and increases, passing through the isotropic
value again at an intermediate lattice, reaches a maximum value and finally
decreases to the FCC value.
The isotropic value of the intermediate lattice is attributed to the
symmetry of the dipole interaction tensor.

Our present theory is of microscopic origin, 
in the sense that we have computed the lattice summation by the 
Ewald-Kornfeld formulation. 
We have not invoked any macroscopic concepts like the Lorentz cavity 
field \cite{Bottcher1,Bottcher2,Lorentz} in the calculations.
However, to corroborate with these established concepts can lead to a 
modification of the Clausius-Mossotti equation.

More precisely, we use the result of the local field to evaluate the 
effective polarizability $\alpha_{\rm eff}$ of the dipole lattice. 
The total field acting on a dipole is the sum of the applied field $E_0$ 
and the local field due to all other dipoles, hence
$$
p = \alpha (E_0 + \beta P),
$$
where $\alpha$ is the polarizability of an isolated dipole and 
$\beta=E/P$ 
is the local field factor. 
We will use $\beta_z$ and $\beta_{xy}$ to denote the local field factors
parallel and perpendicular to the uniaxial anisotropic axis.
Note that $\beta_z=\beta_{xy}=4\pi/3$ when $q=1$.
Let $P=p/V_c$, the above equation becomes a self-consistent equation. 
Solving yields
\begin{eqnarray}
p=\left({\alpha\over {1-\alpha\beta/V_c}}\right) E_0
  \equiv \alpha_{\rm eff} E_0.
\end{eqnarray}
The effective dielectric constant $\epsilon_{\rm eff}$ is given by 
$1+4\pi\alpha_{\rm eff}/V_c$. Thus,
\begin{eqnarray}
{\epsilon_{\rm eff} - 1\over \beta'\epsilon_{\rm eff} + (3-\beta')} 
  ={4\pi \alpha \over 3V'_c},
\end{eqnarray}
where $\beta'=3\beta/4\pi$ and $V'_c=V_c/2$.
For the BCC lattice, $\beta'=1$, $\epsilon_{\rm eff}$ 
satisfies the well-known Clausius-Mossotti equation.
Thus Eq.(10) represents a generalization of the Clausius-Mossotti equation
to the BCT lattices.

\section{Structure transformation via rotating electric fields}

The results of the local field strength as well as the generalized 
Clausius-Mossotti equation allow us to compute the dipole interaction 
energy per particle in the ER solid, similar to calculations of Tao and 
coworkers \cite{Tao}. 
If an uniaxial field ${\bf E}=\hat{\bf z}E_z$ is applied, 
the dipole interaction energy per particle is given by 
$u=-{\bf p}\cdot {\bf E}/2\epsilon_2$, where $\epsilon_2$ is the 
dielectric constant of the host medium. 
Since $E_z=\beta_z (2p/V_c)$ and $p=\alpha E_z$, we obtain
\begin{equation}
u=-{\beta_z \over (V_c/a^3)}\ \ \ {\rm in\ units\ of\ }
 {p^2\over \epsilon_2 a^3}. 
\end{equation}
As $q$ increases from the BCT ground state to the FCC structures, 
the volume of the unit cell increases initially, reaching a maximum at 
the BCC structure, then decreases to the FCC structure. 
As the local electric field strength remains almost constant in the range
$1 < q < 1.5$, the increase in the magnitude of the dipole energy per 
particle is attributed to the decrease in the volume of the unit cell.
Concomitantly, as shown in Fig.3, 
the energy per particle initially increases from the BCT ground state, 
reaching a maximum near the BCC structure and then decreases 
all the way towards the FCC structure. The transformation involves climbing
up an energy barrier beyond which the FCC structure becomes stable, which is
in contrast with the smooth and monotonic transition proposed in previous 
work \cite{Tao98,Sheng}. 

Next we apply a rotating electric field $E_{xy}=rE_z$ in the plane 
perpendicular to the uniaxial electric field, where $r=E_{xy}/E_z$
is the ratio of the rotating to axial field strength. 
The instantaneous electric fields are $E_x=E_{xy} \cos (\omega t+\phi)$ and
$E_y=E_{xy} \sin (\omega t+\phi)$, where $\omega$ is the angular velocity
of the rotating field and $\phi$ is an arbitrary phase angle.
In this case, the dipole interaction energy per particle is modified to:
\begin{equation}
u=-{\beta_z + r^2 \beta_{xy} \over (1+r^2) (V_c/a^3)}\ \ \ 
  {\rm in\ units\ of\ }{p^2\over \epsilon_2 a^3}.
\end{equation}
As shown in Fig.3, when we increase the ratio $r$, the BCT ground state
energy increases while the FCC energy remains unchanged, but there is 
still an energy barrier between the BCT and FCC states for small $r$. 
When $r>2.6$, 
the energy barrier disappears and the FCC structure is the most stable.
Physically one has to break the chains to make the structure 
transformation possible when the strength of the rotating electric field 
is sufficiently large.

From the energy consideration, we suggest that the structure transformation 
be realized in experiments by applying a rotating electric field 
in the plane perpendicular to the uniaxial electric field. 
In this field configuration, both the time average value 
of the induced dipole moment and that of the rotating electric field 
vanish in the plane.
However, the instantaneous dipole moment will induce an overall attractive
force between the particles in the plane perpendicular to the uniaxial
field. This is very different from the previous transformation proposed 
by Tao et al \cite{Tao98} and experimentally verified by Sheng and coworkers 
\cite{Sheng} by using crossed electric and magnetic fields on 
microparticles which possess permanent magnetic dipole moments. 
In our case, the field configuration is all electrical; no magnetic field 
and/or magnetic materials need to be used.
Alternatively, we may apply the same rotating electric field configuration 
to a magnetorheological fluid to achieve the structure transformation.
  
\section{Discussion and conclusion}

Here a few comments on our results are in order.
As mentioned in Ref\cite{Tao98}, the energy difference between the FCC
and hexagonal close-packed (HCP) structures is very small, and is comparable
to the thermal energy, there is a competition between these structures.
Therefore, we will likely find an FCC-HCP mixed structure.
However, it appears that our field configuration helps a four-fold symmetry
in the plane of the rotating field and the FCC structures may be more 
favorable. 
Nevertheless, we are awaiting experimental evidence on the proposed 
structure transformation.

Our calculations have been performed up to the point-dipole 
approximation, 
which is in the same spirit as done by Tao et al \cite{Tao}. 
A dipole-induced-dipole (DID) model, which takes into account the mutual 
polarization effect between touching particles, can drastically improve 
the accuracy towards the fully multipolar calculations \cite{Yu2000}.
Since the DID contribution becomes important for small reduced separation
$\sigma=d/2R <1.1$ \cite{Yu2000}, we can simplify the calculations by 
limiting ourselves to touching spheres only. We may omit the nontouching 
spheres (unless the spheres get very close so that $\sigma=d/2R <1.1$). 
Each sphere is in contact with 8 neighboring spheres in the BCT lattices
and 12 neighboring spheres in the FCC lattice and the number of DID 
images dipoles is therefore finite.
The generalization thus includes the original point dipoles as well as 
all the DID images dipoles at well-defined positions in the unit cell and 
the general formula [Eq.(4)] can indeed be used.
The incorporation of the more accurate DID model into the Ewald-Kornfeld 
formulation is underway. 
However, we believe that the point-dipole results remain qualitatively 
correct.

In conclusion, we have applied the Ewald-Kornfeld formulation to a 
BCT lattice of point dipoles to examine the effects of structure 
transformation on the local field distribution.
We have found that the local field exhibits a non-monotonic transition 
from BCT to FCC.
Moreover, we showed that the change in the local field can lead to a 
generalized Clausius-Mossotti equation for the BCT lattices.

\section*{Acknowledgments}
This work was supported by the Research Grants Council of the Hong Kong 
SAR Government under grant CUHK 4284/00P.
K.W.Y. acknowledges the hospitality received during his participation in
the Workshop on Soft Matter, hosted by the Institute of Physics of the 
Chinese Academy of Sciences, where the present work was initiated.
We acknowledge discussion with Jones T. K. Wan on various aspects of 
the project.

\begin{figure}[h]
\caption{A sequence of unit cells of BCT lattices during structure 
transformation:
(a) Tao's BCT lattice, (b) BCC lattice, (c) intermediate lattice and
(d) FCC lattice.}
\end{figure}
 
\begin{figure}[h]
\caption{(a) The local field factor plotted against $\log_{10} q$ for 
dipole moments along the uniaxial anisotropic axis. 
(b) Similar to (a), but for dipole moments perpendicular to the uniaxial
anisotropic axis.
(c) and (d): Magnified versions of (a) and (b) respectively to show the 
different lattices during structure transformation.}
\end{figure}

\begin{figure}[h]
\caption{The dipole interaction energy per particle plotted against $q$
 for different ratio $r=E_{xy}/E_z$ of the rotating to axial field strength. 
 For hard spheres, the accessible regions are within the vertical lines
 depicted by Tao's BCT and FCC. Along Tao's BCT line and from bottom
 to top, $r=$ 0.0, 1.0, 1.6, 2.0, 2.2, 2.4, 2.6, 3.0, 4.0 and $\infty$.
 It is evident that the FCC structure is the most stable for large $r$.}
\end{figure}

\newpage
\centerline{\epsfig{file=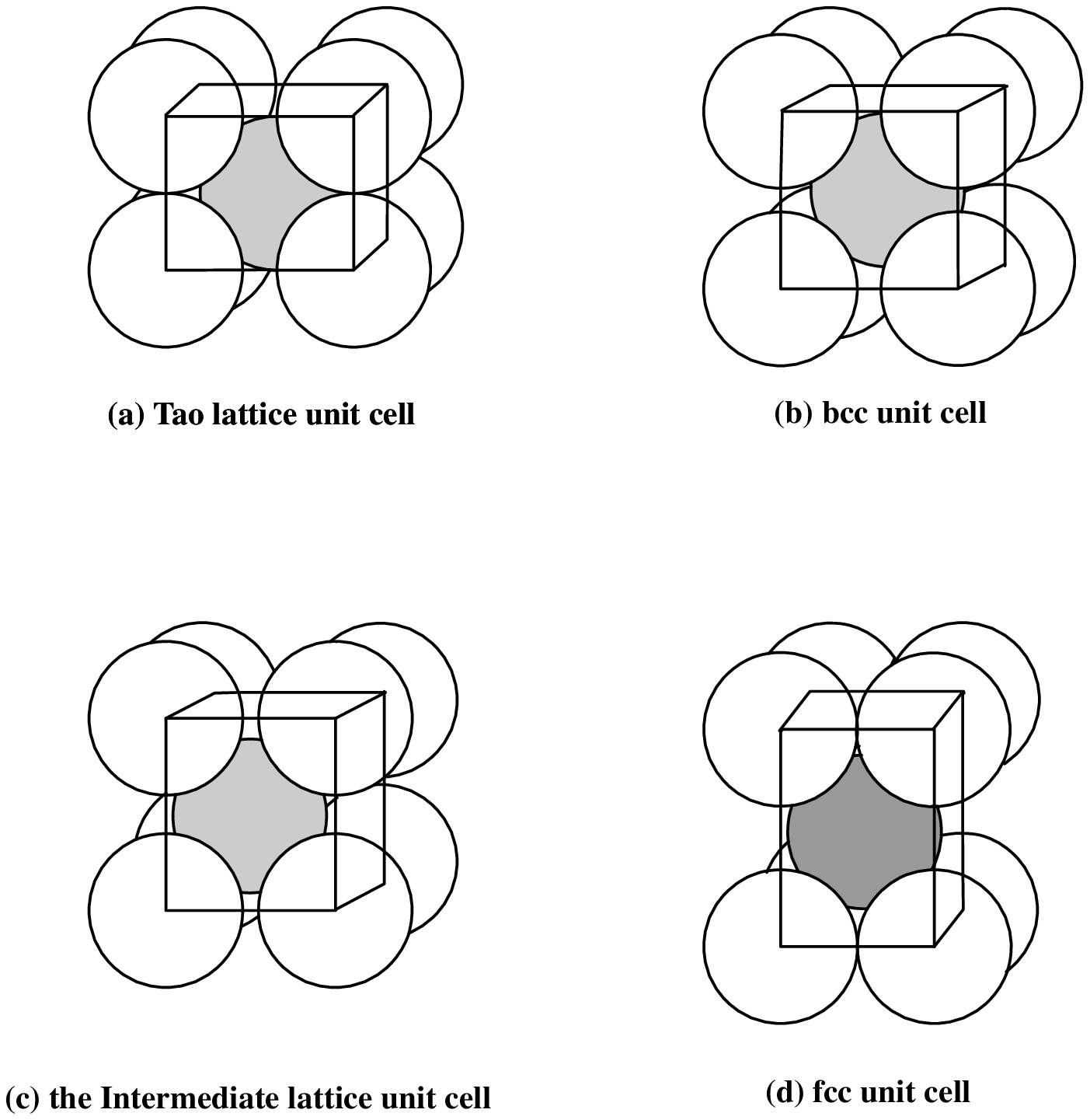,width=\linewidth}}
\centerline{Fig.1/Lo and Yu}

\newpage
\centerline{\epsfig{file=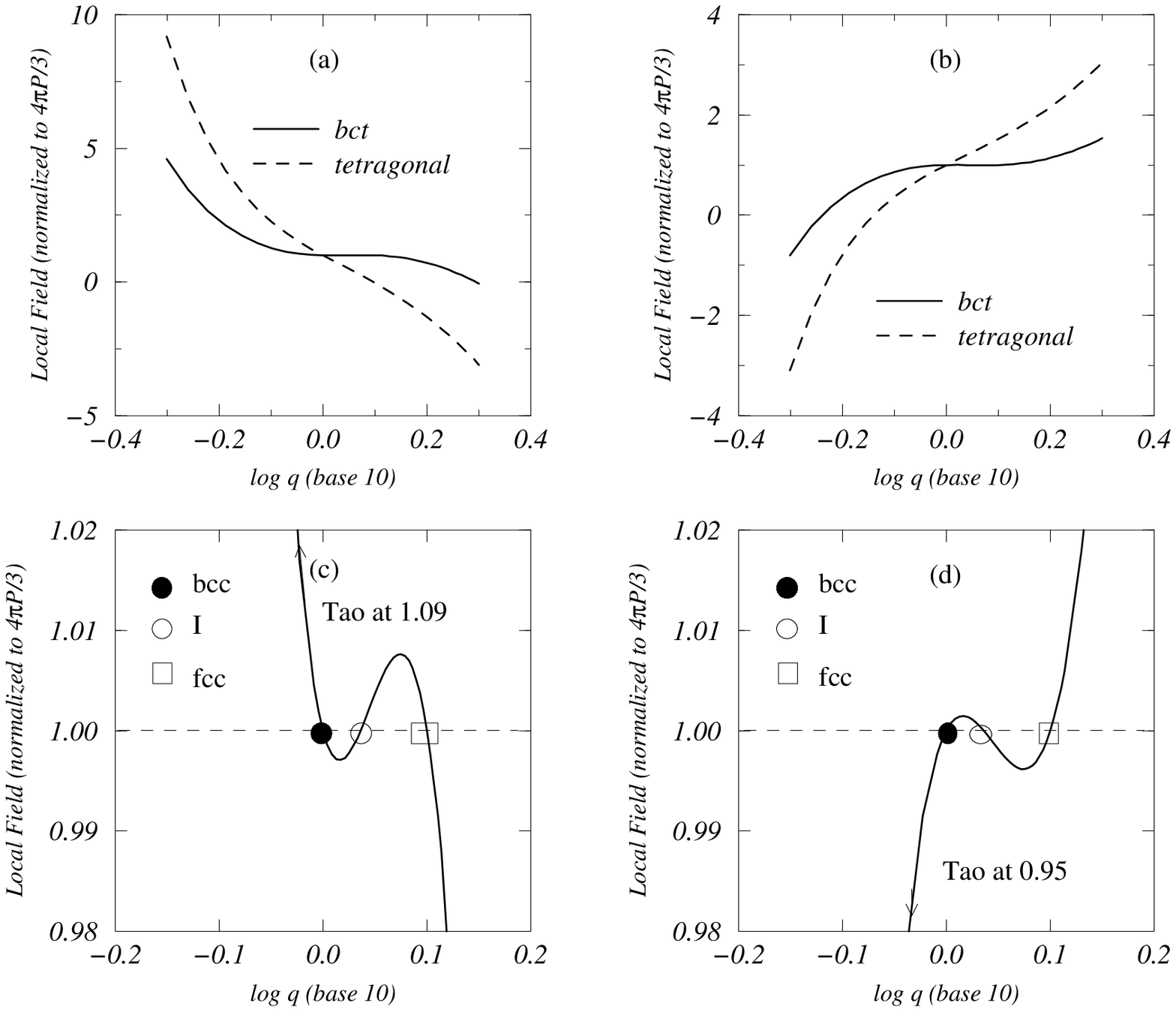,width=\linewidth}}
\centerline{Fig.2/Lo and Yu}

\newpage
\centerline{\epsfig{file=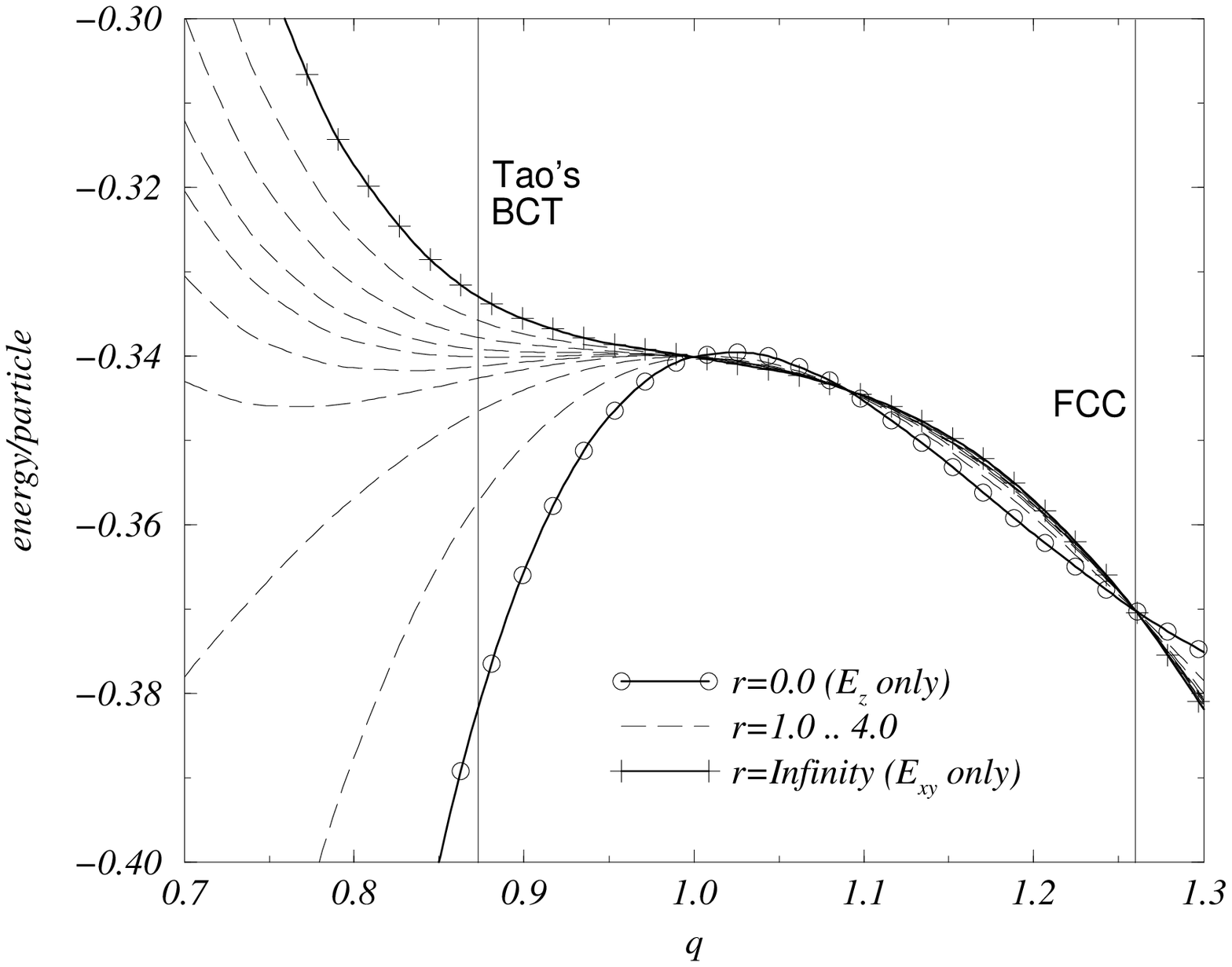,width=\linewidth}}
\centerline{Fig.3/Lo and Yu}

\end{document}